# Superconductivity in layered $CeO_{0.5}F_{0.5}BiS_2$

Rajveer Jha and V.P.S. Awana

Quantum Phenomena and Application Division, National Physical Laboratory (CSIR), Dr. K. S. Krishnan Road, New Delhi 110012, India

We report appearance of superconductivity in $CeO_{0.5}F_{0.5}BiS_2$. The bulk polycrystalline samples $CeOBiS_2$ and $CeO_{0.5}F_{0.5}BiS_2$ are synthesized by conventional solid state reaction route via vacuum encapsulation technique. Detailed structural analysis showed that the studied $CeO_{0.5}F_{0.5}BiS_2$ compound is crystallized in tetragonal P4/nmm space group with lattice parameters a = 4.016(3) Å, c = 13.604(2) Å. DC magnetization measurement (MT-curve) shows the ferromagnetic signal at the low temperature region. The superconductivity is established in $CeO_{0.5}F_{0.5}BiS_2$ at $T_c^{onset}$ = 2.5K by electrical transport measurement. Under applied magnetic field both $T_c$ onset and $T_c$ ($\rho$ =0) decrease to lower temperatures and an upper critical field [$H_{c2}(0)$] above 1.2Tesla is estimated. The results suggest coexistence of ferromagnetism and superconductivity for the $CeO_{0.5}F_{0.5}BiS_2$ sample.

Key Words: *$BiS_2$ based new superconductor, structure, ferromagnetic and transport properties.*

PACS number(s): 74.10.+v, 74.70.-b, 74.70. Dd, 75.50.Gg, 74.25.Fy

*Corresponding Author

Dr. V. P. S. Awana, Principal Scientist

E-mail: awana@mail.npindia.org

Ph. +91-11-45609357, Fax-+91-11-45609310

Homepage www.fteewebs.com/vpsawana/

Introduction:

The recent discovery of the $BiS_2$-based layered superconductors $Bi_4O_4S_3$ [1, 2] and $REO_{0.5}F_{0.5}BiS_2$ (RE-La, Nd, Pr, Yb & Ce) [3-9] had been of tremendous interest for the condensed matter community. These compounds are very similar in crystal structure to the layered high $T_c$ cuprate and Fe-based superconductors. In the high $T_c$ cuprate superconductors, the superconducting $CuO_2$ layers are separated by blocking layers [10], while in case of Fe-based compounds, the same role is played by the FeAs layers [11]. The new class of superconductors possesses superconducting $BiS_2$ layers being separated by charge reservoir blocking layers [12, 13]. Both experimental and theoretical studies showed that $BiS_2$ layered compounds exhibit multiband behaviors with dominant electron charge carriers originating from the Bi $6p_x$ and $6p_y$ bands in the normal state [1, 14]. Theoretical studies especially for $LaO_{0.5}F_{0.5}BiS_2$ compound indicated that the Fermi level crosses conduction bands and thus yields electron pockets [15]. There is also a suggestion that $BiS_2$ based new superconductors could be conventional superconductors with strong electron phonon coupling [16-18]. The structural instability places these compounds in close proximity to the competing ferroelectric and charge density wave (CDW) phases [15]. This is similar to that as the spin-density wave (SDW) instability in iron based superconductors [19]. The experimental studies for $BiS_2$-based materials have focused on increasing the charge carrier concentration via chemical substitution within the blocking layer as well as through a reduction of the unit cell volume via the application of an external pressure, for example, recent studies on the $BiS_2$-based compounds involving chemical substitution [20, 21]. The $LaOBiS_2$ and $NdOBiS_2$, having analogous $BiS_2$ layers, had been found to show superconductivity upon electron doping [22]. It has been proposed that superconductivity and ferromagnetism coexist in the $CeO_{1-x}F_xBiS_2$ compounds; the ferromagnetic order state presumably arises due to the local moments of Ce at low temperatures [9]. In this short communication we report superconductivity of $CeO_{0.5}F_{0.5}BiS_2$ with a typical $BiS_2$ layer. Both compounds $CeOBiS_2$ and $CeO_{0.5}F_{0.5}BiS_2$ crystallized in the tetragonal *P*4/*nmm* space group. We found the $CeO_{0.5}F_{0.5}BiS_2$ compound shows ferromagnetic transition near 4K in magnetization measurements. It is found that the parent phase is a bad metal, and F doped $CeOBiS_2$ compound exhibits superconductivity at 2.5K along with 4K ferromagnetism.

Experimental:

Bulk polycrystalline samples $CeOBiS_2$ and $CeO_{0.5}F_{0.5}BiS_2$ were synthesized by standard solid state reaction route via vacuum encapsulation. High purity Ce, Bi, S, $CeF_3$, and $CeO_2$ are weighed in stoichiometric ratio and ground thoroughly in a glove box under high purity argon atmosphere. The mixed powders are subsequently palletized and vacuum-sealed ($10^{-3}$ Torr) in a quartz tube. Sealed quartz ampoule is placed in tube furnace and heat treated at $700^0C$ for 12h with the typical heating rate of $2^oC$/min., and subsequently cooled down slowly over a span of six hours to room temperature. This process was repeated twice. X-ray diffraction (*XRD*) was performed at room temperature in the scattering angular (*2θ*) range of $10^o$-$80^o$ in equal *2θ* step of $0.02^o$ using *Rigaku Diffractometer* with *Cu K$_α$* ($\lambda$ = 1.54Å). Rietveld analysis was performed using the standard *FullProf* program. The electrical transport and magnetization measurements were performed on Physical Property Measurements System (*PPMS*-14T, *Quantum Design*) as a function of both temperature and applied magnetic field.

Results and Discussion:

XRD data at room temperature of $CeOBiS_2$ and $CeO_{0.5}F_{0.5}BiS_2$ samples are shown in Figure 1. Rietveld refinement of XRD patterns is carried out for the both compounds. These compounds are crystallized in tetragonal structure in space group P4/nmm with small impurity peaks of Bi and $Bi_2S_3$. Rietveld refined lattice parameters, atomic coordinates, and site occupancy are shown in the Table I. The Rietveld fitted results exhibited that the lattice parameter *a* axis of $CeOBiS_2$ is 4.016Å and the same is increased for $CeO_{0.5}F_{0.5}BiS_2$ sample to 4.037 Å. The *c*-axis lattice constant is 13.604 Å and 13.407 Å respectively for $CeOBiS_2$ and $CeO_{0.5}F_{0.5}BiS_2$ samples. The decrease in c-axis parameter indicates that F is doped successfully at the O site as the ionic radius of F is smaller than that of O. These results are in good agreement with the reported phase diagrams for $NdO_{1-x}F_xBiS_2$ and $CeO_{1-x}F_xBiS_2$ compounds [4, 8].

Figure 2 depict temperature dependence of magnetic susceptibility at 10Oe applied field for the superconducting $CeO_{0.5}F_{0.5}BiS_2$ sample, we observed a strong ferromagnetic signal below about 4 K. No diamagnetic signal is observed down to 1.9 K. Perhaps possible lower temperature (below 2.4K, seen in transport measurements, next section) superconducting diamagnetism is prevailed over by ferromagnetism. It seems the Ce sub-lattice is magnetically ordered at below about 4K. Inset of Fig.2 shows magnetization-hysteresis (MH) curve at 2K, indicating clearly the ferromagnetic nature. Similar magnetization results are obtained for studied $CeOBiS_2$ as well. This result is in confirmation with an earlier report on $CeO_{1-x}F_xBiS_2$ compounds [8]. May it be that that superconductivity occurs in the $BiS_2$ layers along with ferromagnetic order in the CeO layer.

Figure 3(a) presents the resistivity versus temperature ($\rho$–T) plots for the $CeO_{0.5}F_{0.5}BiS_2$ sample in the applied magnetic field of up to 10 kOe in temperature range of 1.9 – 3K. The result is similar to that as reported for $LaO_{0.5}F_{0.5}BiS_2$ superconductor. The superconductivity sets at the onset of $\rho$–T plots and is complete at $\rho = 0$. In this case, with the application of the magnetic field, both the onset and offset $T_c$ shift toward lower temperature. The inset of the same shows the resistivity in expanded temperature range of 1.9–250K of the $CeOBiS_2$ and $CeO_{0.5}F_{0.5}BiS_2$. The parent compound is showing semiconducting like behavior from 250K down to 1.9K. The superconducting $T_c$ onset is observed for $CeO_{0.5}F_{0.5}BiS_2$ compound in resistivity curve at 2.5K and the zero resistivity $T_c(\rho = 0)$ at 1.9K.

Figure 3 (b) shows the temperature dependence upper critical field $\mu_0H_{c2}(T)$, which has been fitted through Ginzburg-Landau theory by using $\rho_n = 90\%$ criterion. In the studied temperature range the dependence of field $\mu_0H_{c2}(T)$ is nearly linear. The upper critical field $H_{c2}$ evolves with temperature from the formula $H_{c2}(T) = H_{c2}(0)(1 - t^2)/(1 + t^2)$ where $t = T/T_c$. From the fitting, we can clearly see that, initially the behavior of $H_{c2}$ with T is linear near $T_c$ and extends up to a temperature of 0.5K and after that the same nearly saturates. The $H_{c2}(0)$ for the sample is found to be about 1.2 Tesla. The $H_{c2}(0)$ value determined by us is in agreement with other reported literature for $BiS_2$ based superconducting compounds [6,8].

## Conclusion:

In conclusion, we have successfully synthesized $BiS_2$-based $CeO_{0.5}F_{0.5}BiS_2$ superconductor. MT-curve showed ferromagnetic signal at below 4K. $CeOBiS_2$ sample showed bad metal behavior and $T_c = 2.5K$ has been observed in $CeO_{0.5}F_{0.5}BiS_2$ sample. R(T)H of $CeO_{0.5}F_{0.5}BiS_2$ exhibited an upper critical field of above 1.2Tesla. We have observed coexistence of ferromagnetic and superconductivity for the $CeO_{0.5}F_{0.5}BiS_2$ compound. Our results are in confirmation with only report of superconductivity in polycrystalline $CeO_{0.5}F_{0.5}BiS_2$ [8]. Very recently, after submission of this work, an article has appeared on cond-mat arxiv related to single crystal growth of superconducting $CeO_{1-x}F_xBiS_2$ with magnetic anomalies in low temperature regime [23].

Acknowledgements

The authors are grateful for the encouragement and support from Director NPL for this work. Rajveer Jha would like to thank the CSIR for providing the SRF scholarship to pursue his Ph.D. This work is also financially supported by DAE-SRC outstanding investigator award scheme on search for new superconductors.

Reference:

1. Y. Mizuguchi, H. Fujihisa, Y. Gotoh, K. Suzuki, H. Usui, K. Kuroki, S. Demura, Y. Takano, H. Izawa, O. Miura Phys. Rev. B, **86**, 214518 (2012).

2. S. K. Singh, A. Kumar, B. Gahtori, Shruti; G. Sharma, S. Patnaik, V. P. S. Awana, J. Am. Chem. Soc., **134**, 16504 (2012).

3. Y. Mizuguchi, S. Demura, K. Deguchi, Y. Takano, H. Fujihisa, Y. Gotoh, H. Izawa, O. Miura, J. Phys. Soc. Jpn., **81**, 114725 (2012).

4. S. Demura, Y. Mizuguchi, K. Deguchi, H. Okazaki, H. Hara, T. Watanabe, S. J. Denholme, M. Fujioka, T. Ozaki, H. Fujihisa, Y. Gotoh, O. Miura, T. Yamaguchi, H. Takeya, and Y. Takano, J. Phys. Soc. Jpn., **82** 033708 (2013).

5. R. Jha, A. Kumar, S. K. Singh, V.P.S. Awana, J Supercond Nov Magn **26**, 499 (2013).

6. V.P.S. Awana, A. Kumar, R. Jha, S. K. Singh, A. Pal, Shruti, J. Saha, S. Patnaik, Solid State Communications **157**, 21 (2013).

7. R. Jha, A. Kumar, S. K. Singh, and V. P. S. Awana, J. App. Phy. **113**, 056102 (2013).

8. Jie Xing, Sheng Li, Xiaxin Ding, Huan Yang, and Hai-Hu Wen, Phys. Rev. B **86**, 214518 (2012).

9. D. Yazici, K. Huang, B.D. White, A.H. Chang, A.J. Friedman and M.B. Maple, Philosophical Magazine **93**, 673 (2013).

10. J. G. Bednorz and K.A. Muller, Z. Phys. B **64**, 189 (1986).

11. Y. Kamihara, T. Watanabe, M. Hirano, H. Hosono, J. Am. Chem. Soc. **130**, 3296 (2008).

12. H. Kotegawa, Y. Tomita, H. Tou, H. Izawa, Y. Mizuguchi, O. Miura, S. Demura, K. Deguchi, and Y. Takano, J. Phys. Soc. Jpn., **81** 103702 (2012).

13. B. Li, Z. W. Xing and G. Q. Huang, Euro Phys. Lett., **10**, 47002 (2013).

14. H. Usui, K. Suzuki, and K. Kuroki, Phys. Rev. B **86**, 220501(R) (2012).

15. X. Wan, H. C. Ding, S. Y. Savrasov, and C. G. Duan, Phys. Rev. B **87**, 115124 (2013).

16. G. B. Martins, A. Moreo, and E. Dagotto, Phys. Rev. B **87**, 081102(R) (2013).

17. I. R. Shein, A. L. Ivanovskii, . *Teoret. Fiz.* **96**, 859 (2012).

18. T. Yildirim, Phys. Rev. B **87**, 020506(R) (2013).

19. K. Deguchi, Y. Mizuguchi, S. Demura, H. Hara, T. Watanabe, S. J. Denholme, M. Fujioka, H. Okazaki, T. Ozaki, H. Takeya, T. Yamaguchi, O. Miura and Y. Takano, Euro. Phys. Lett., **101** 17004 (2013).

20. S.G. Tan, L. J. Li, Y. Liu, P. Tong, B.C. Zhao, W.J. Lu, Y.P. Sun, Physica C **483**, 94 (2012).

21. Corentin Morice, Emilio Artacho, y Sian E. Dutton, Daniel Molnar, Hyeong-Jin Kim, and Siddharth S. Saxena, arXiv:1305.1201v1 (2013).

22. Y. Mizuguchi, H. Fujihisa, Y. Gotoh, K. Suzuki, H. Usui, K. Kuroki, S. Demura, Y. Takano, H. Izawa, and O. Miura, Phys. Rev. B **86**, 220510(R) (2012).

23. M. Nagao, A. Miura, S. Demura, K. Deguchi, S. Watauchi, T.Takei, Y. Takano, N. Kumada and I. Tanaka, arXiv:1310.1213.

Table 1 Atomic coordinates, Wyckoff positions, and site occupancy for studied $CeO_{0.5}F_{0.5}BiS_2$.

| Atom | x | y | z | site | Occupancy |
|------|------|------|----------|------|-----------|
| Ce   | 0.2500 | 0.2500 | 0.099(3) | 2c | 1 |
| Bi   | 0.2500 | 0.2500 | 0.625(5) | 2c | 1 |
| S1   | 0.2500 | 0.2500 | 0.378(1) | 2c | 1 |
| S2   | 0.2500 | 0.2500 | 0.811(2) | 2c | 1 |
| O/F  | 0.7500 | 0.2500 | 0.000    | 2a | 0.5/0.5 |

**Figure Captions:**

Figure 1 (a): Observed (*Red circles*) and calculated (*solid lines*) XRD pattern of $CeOBiS_2$ and $CeO_{0.5}F_{0.5}BiS_2$ compounds at room temperature.

Figure 2: Temperature dependence of the dc magnetization of the $CeO_{0.5}F_{0.5}BiS_2$ compound. Inset of the figure shows Isothermal MH curve at 2 K of the sample $CeO_{0.5}F_{0.5}BiS_2$.

Figure 3: (a) Resistivity vs. temperature ($\rho$ – T) behavior of $CeO_{0.5}F_{0.5}BiS_2$ with and without applied fields. Inset of the figure shows the $\rho$ – T for $CeOBiS_2$ and $CeO_{0.5}F_{0.5}BiS_2$ in the temperature range 250 K to 1.9 K. (b) Temperature dependent upper critical field ($H_{c2}$) fitted by the GL equation for the $\rho_n$ =90% of $CeO_{0.5}F_{0.5}BiS_2$ sample.

Figure 1

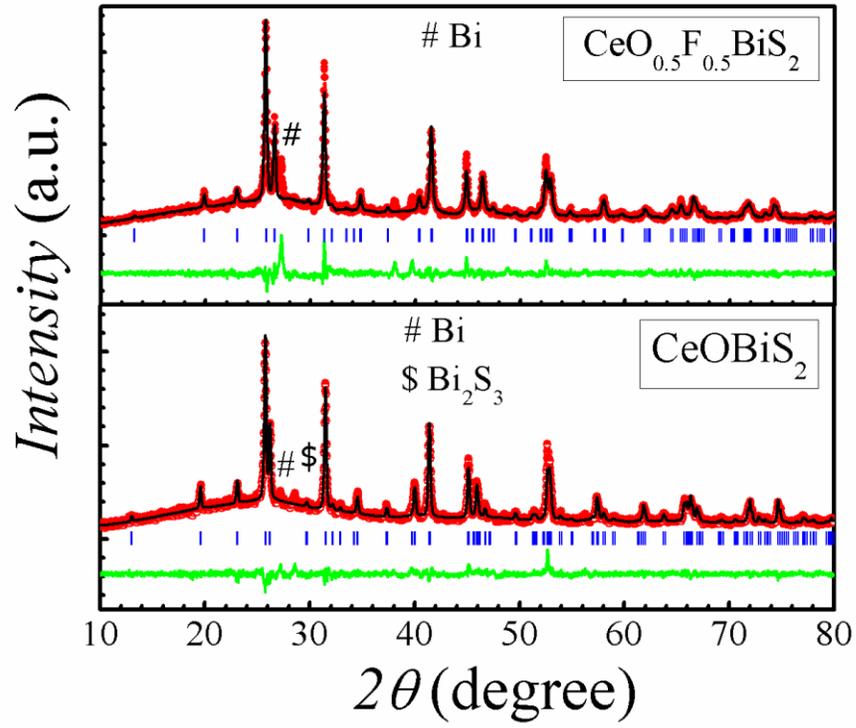

Figure 2

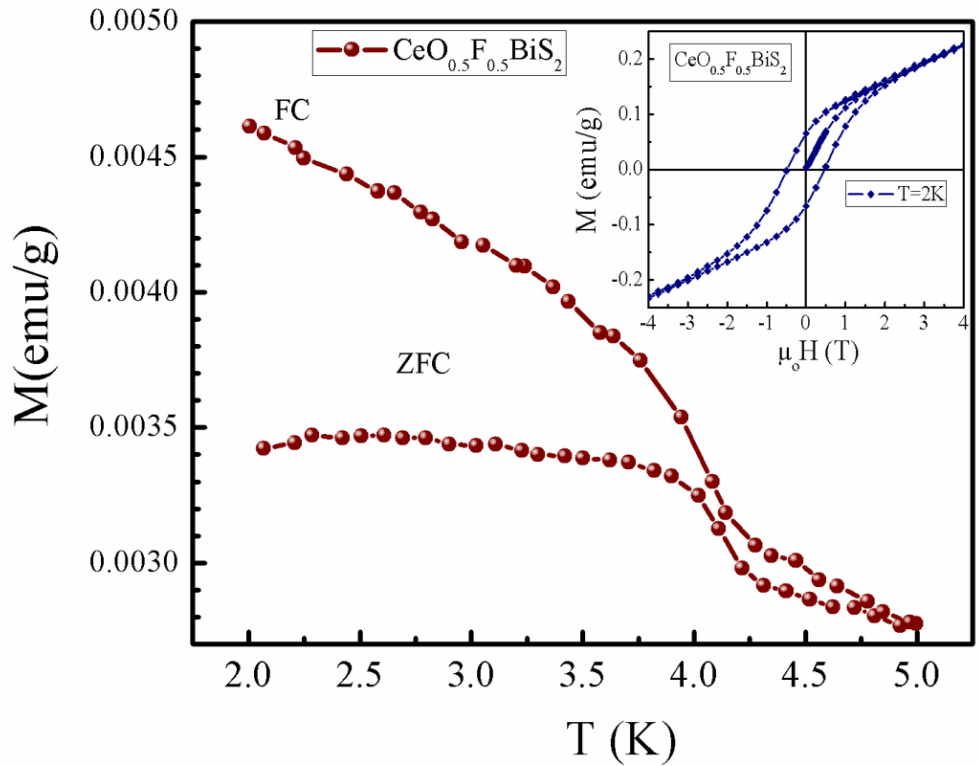

Figure 3(a):

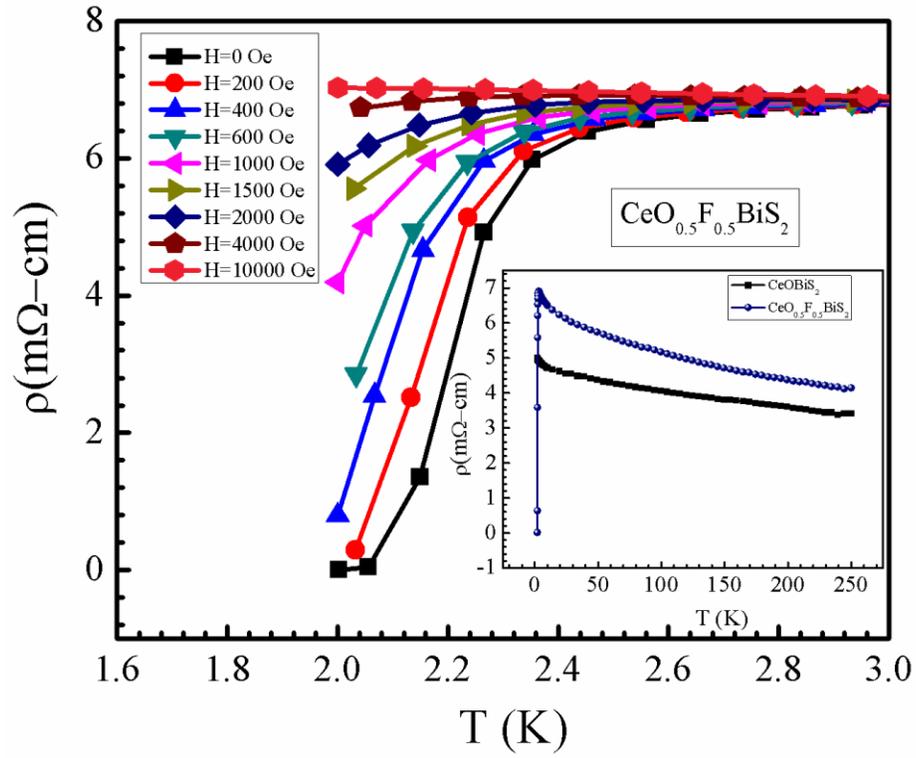

Figure 3(b):

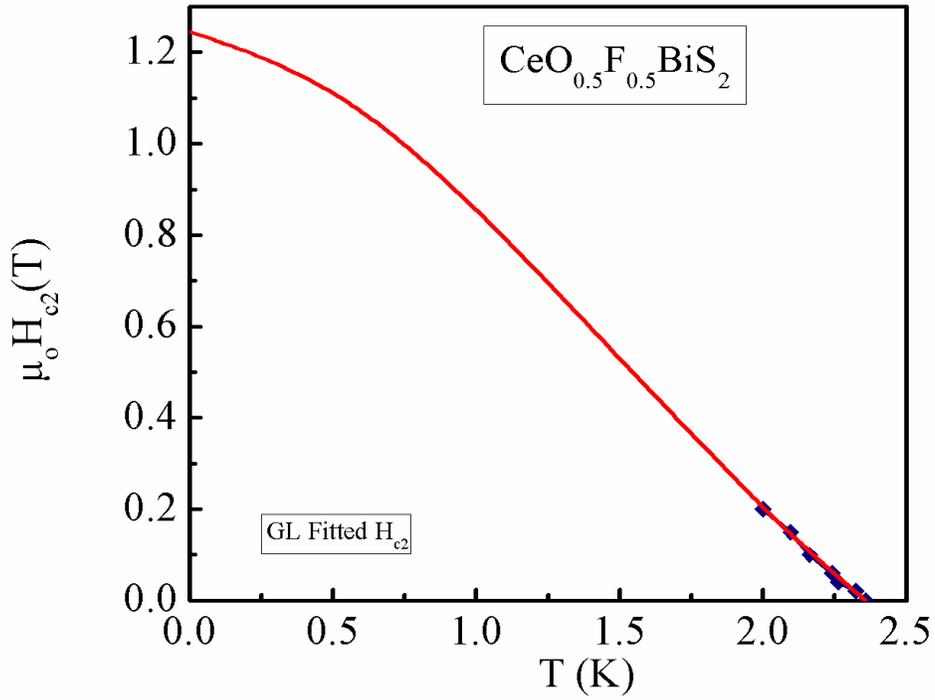